\documentclass[conference]{IEEEtran}
\IEEEoverridecommandlockouts

\usepackage[utf8]{inputenc}
\usepackage[T1]{fontenc}
\usepackage{cite}
\usepackage{amsmath,amssymb,amsfonts}
\usepackage{algorithmic}
\usepackage{graphicx}
\usepackage{textcomp}
\usepackage{xcolor}
\usepackage{booktabs}
\usepackage{array}
\usepackage{url}
\usepackage{tikz}
\usetikzlibrary{arrows.meta,positioning,shapes,fit,backgrounds,calc}
\usepackage[hidelinks]{hyperref}
\usepackage{xspace}
\usepackage{float}
\usepackage{flushend}

\def\BibTeX{{\rm B\kern-.05em{\sc i\kern-.025em b}\kern-.08em
    T\kern-.1667em\lower.7ex\hbox{E}\kern-.125emX}}

\newcommand{\sys}{CYBER-BRIDGE\xspace}
\newcommand{\ex}{\normalfont\scriptsize\itshape\color{black!55}}

\title{The \sys project: An AI-Powered Platform for
       Cross-Border Cybersecurity, Compliance
       and Forensics in the EU}

\author{%
\IEEEauthorblockN{%
N.~Kekatos\textsuperscript{1},
G.~Koutidis\textsuperscript{1},
C.~Bratsas\textsuperscript{2},
S.~Bucur\textsuperscript{3},
A.-M.~Chirlejan\textsuperscript{4},
M.~Curca\textsuperscript{4},
A.~Fountoulakis\textsuperscript{5},\\
M.~Ioannou\textsuperscript{6},
P.~Lacatus\textsuperscript{7},
A.~Lekidis\textsuperscript{8},
E.~Charalambous\textsuperscript{9},
A.~Michael\textsuperscript{9},
T.~Nianios\textsuperscript{1},
S.~Purcaru\textsuperscript{4},
A.~Voulgaridis\textsuperscript{10}}
\IEEEauthorblockA{%
\begin{minipage}{0.98\textwidth}\centering
\footnotesize
\textsuperscript{1}Clone Systems, CY
~--~
\texttt{\{nkekatos,\,gkoutidis,\,tnianios\}@clone-systems.com}\quad
\textsuperscript{2}IHU, GR
~--~
\texttt{bratsas@ihu.gr}\quad
\textsuperscript{3}DGPI, RO
~--~
\texttt{bucur.stefan@dgpi.ro}\quad
\textsuperscript{4}DNSC, RO
~--~
\texttt{\{Mihaela.Curca,\,alexandra.chirlejan,\,simona.purcaru\}@dnsc.ro}\quad
\textsuperscript{5}Eximprod Eng., RO
~--~
\texttt{angelos.fountoulakis@epg.ro}\quad
\textsuperscript{6}Boltonshield, HU
~--~
\texttt{michael@boltonshield.com}\quad
\textsuperscript{7}I-ENERGYLINK, RO
~--~
\texttt{paul.lacatus@i-energylink.eu}\quad
\textsuperscript{8}Univ.\ of Thessaly, GR
~--~
\texttt{alekidis@uth.gr}\quad
\textsuperscript{9}Elias Neocleous \& Co.\ LLC, CY
~--~
\texttt{\{emilios.charalambous,\,Avgi.michael\}@neo.law}\quad
\textsuperscript{10}CERTH, GR
~--~
\texttt{antonismv@iti.gr}
\end{minipage}
}}
\begin{document}
\maketitle
{\renewcommand{\thefootnote}{}\footnotetext{\copyright~2026 IEEE. Personal use of this material is permitted. Permission from IEEE must be obtained for all other uses, in any current or future media, including reprinting/republishing this material for advertising or promotional purposes, creating new collective works, for resale or redistribution to servers or lists, or reuse of any copyrighted component of this work in other works. This work has been accepted for publication at the 2026 IEEE International Conference on Cyber Security and Resilience (IEEE CSR 2026).}}

\begin{abstract}
Cross-border cyber attacks on European SMEs and
critical-infrastructure operators expose a structural cyber-resilience
gap: cybersecurity (threat sharing, detection, response), compliance
(framework assessment, certification) and digital forensics
(cross-jurisdictional evidence) are jointly regulated by
NIS2, the CRA, GDPR and the LED, yet their tooling
evolved in silos and rarely interoperates across Member States.
The EU-funded \sys project integrates the three domains under one
architecture and a hash-chained evidence layer: AI-driven CTI sharing
with ENISA-aligned notification, multi-framework assessments with
retrieval-augmented advisory, and RFC~3161-anchored custody with
lawful GDPR/LED exchange, all aligned with MITRE
ATT\&CK. A sanitised replica of the back-end image
hosts a Cyber-Range for upskilling. The contribution is not the
individual tools but the reconciliation mechanism that lets one
incident produce, from a single evidence record, a cross-border CTI
advisory, the regulatory notifications its deadlines demand and a
re-verifiable forensic pack; we set out why binding three differently
regulated evidence regimes onto one record is non-trivial and how the
shared evidence layer resolves the resulting tensions. This paper
introduces the project's scope, requirements, pilots, architecture and
toolkits; quantitative KPIs follow in later deliverables.
\end{abstract}


\section{Introduction}
\label{sec:intro}

European cybersecurity is now unavoidably cross-border. Attacks on
small and medium-sized enterprises (SMEs) and critical-infrastructure
operators routinely span Member States, and effective response demands
coordinated action by security operations centres (SOCs), computer
security incident response teams (CSIRTs) and law-enforcement agencies
across three operational
domains: cybersecurity (threat sharing, detection, response),
compliance (framework assessment, certification) and digital
forensics (cross-jurisdictional evidence). The EU regulatory
perimeter has expanded in step. The principal instruments are the
NIS2 Directive~\cite{nis2directive}, the Cyber Resilience Act
(CRA)~\cite{craregulation}, the General Data Protection Regulation
(GDPR)~\cite{gdpr2016} and the Law Enforcement Directive
(LED)~\cite{led2016}, complemented by Directive
2013/40/EU~\cite{dir2013_40}, the Cybersecurity
Act~\cite{csact2019} and European Union Agency for Cybersecurity
(ENISA) guidance~\cite{enisa2023,enisaincidentnotif2022}. As a result, an
essential entity or product manufacturer placing services on the EU
market must now satisfy several overlapping frameworks with distinct
evidence and notification expectations.

The schedule is tight: NIS2 transposition (October~2024 deadline)
and the CRA's 36-month application window pull essential entities,
manufacturers and external auditors onto a common alignment
timeline. Yet the tooling that supports each domain (scanners,
governance, risk and compliance (GRC) suites, threat-intelligence
platforms and
forensic chains) evolved in silos and rarely interoperates, and
CSIRTs, SOCs and law-enforcement agencies remain equipped for
vertical coverage of one domain, not the horizontal cross-border
evidence flow an end-to-end incident demands.

The EU-funded \sys project addresses this gap. A 36-month effort
co-funded by the European Cybersecurity Competence Centre under the
Digital Europe Programme,\footnote{\url{https://www.cyberbridge-eu.eu/}}
it integrates the three domains under one architecture and a shared
hash-chained evidence layer (Fig.~\ref{fig:platform-overview}):
\textbf{cybersecurity} (cross-border cyber-threat intelligence (CTI)
sharing, scanning,
NIS2-aligned incident response), \textbf{compliance} (CRA, GDPR,
ISO/IEC 27001~\cite{iso27001}, NIS2, NIST CSF~\cite{nistcsf}) and
\textbf{forensics} (RFC~3161-anchored custody~\cite{rfc3161}, legal
hold, lawful GDPR/LED exchange), plus a Cyber-Range on a sanitised
back-end replica for upskilling.

\textbf{Hypothesis.} We posit that binding the three domains to a
single hash-chained evidence record, rather than integrating them
only at the report level, lets one cross-border incident produce its
CTI advisory, its regulatory notifications and a re-verifiable
forensic pack from the same data: an analyst records a finding once
and the platform derives the partner-facing indicator, the
deadline-bound notification and the custody record, removing the
manual re-keying that fragmented tooling imposes. The five pilots
(Section~\ref{sec:reqs:pilots}) test this claim.

\textbf{Contribution.} The contribution is not the individual tools,
each well established, but the reconciliation mechanism that makes one
artefact valid at once as a threat indicator, a compliance evidence
item and a forensic custody record. We make it concrete through
(i) an articulation of why this integration is technically non-trivial
and how the shared evidence layer resolves the resulting tensions
(Section~\ref{sec:arch:integ}); (ii) a threat model mapped to MITRE
ATT\&CK~\cite{mitreattack}, positioning \sys against peer EU efforts
(AI4CYBER~\cite{ai4cyber}, CyberSEAS~\cite{cyberseas},
SECANT~\cite{secant}) and open-source alternatives
(Table~\ref{tab:peers}); and (iii) an end-to-end walk-through
(Section~\ref{sec:tk-cyber}) tracing one attack step from detection
through ATT\&CK mapping, regulatory-clock scheduling and evidence
enrolment, validated through five sectoral pilots under NIS2, CRA,
GDPR and LED.

All operational telemetry (threat indicators, scans, dark-web
findings, incidents and forensic artefacts) is exposed as evidence
linked to framework objectives, regulatory clocks and partner
organisations. The project objectives are:

(1)~\emph{cross-border CTI sharing and incident response} --- AI-driven
exchange across SOCs, CSIRTs and law-enforcement partners over CISA
Known Exploited Vulnerabilities (KEV), the EU Vulnerability Database
(EUVD), MITRE ATT\&CK and the National Vulnerability Database (NVD),
with ENISA-aligned notification;
(2)~\emph{automated regulatory compliance} --- a multi-framework engine
hosting CRA, GDPR, ISO/IEC 27001, NIS2 and NIST CSF as
Framework\,$\to$\,Chapter\,$\to$\,Objective, with snapshots, gap
detection and regulatory-change monitoring;
(3)~\emph{AI Assistant and Compliance Agent} --- a retrieval-augmented
generation (RAG)-grounded assistant and a roadmap-producing agent on a
self-hosted Phi-4;
(4)~\emph{EU-aligned security tooling} --- network, web, static
application security testing (SAST), software composition analysis
(SCA) and software bill of materials (SBOM) scanning, CTI feeds and
policy-gated dark-web probing, with active scanners kept off live
operational technology (OT);
and (5)~\emph{auditable evidence, certification, forensics and
training} --- \texttt{Evidence} items (SHA-256, RFC~3161, legal hold,
framework binding), verifiable certificates, lawful GDPR/LED exchange,
and a Cyber-Range on a sanitised replica.

\textbf{Related work and positioning.}
\label{sec:sota}
\sys builds on Nmap, OWASP ZAP, Semgrep~\cite{semgrep},
OSV~\cite{osvscanner} and Syft~\cite{syft}
(CycloneDX\footnote{\url{https://cyclonedx.org/}}) for CRA
vulnerability-handling and SBOM
obligations. CTI platforms use MITRE ATT\&CK~\cite{mitreattack},
CISA KEV~\cite{cisakev}, NVD~\cite{nvd}/EUVD~\cite{euvd} and
STIX/TAXII~\cite{stixtaxii} with Traffic Light Protocol (TLP) markings
for CTI. GRC
suites~\cite{eramba,simplerisk} rarely pre-map NIS2 or the CRA and
treat security telemetry separately; compliance advisors increasingly
apply large language models (LLMs) via RAG~\cite{lewis2020rag}, and
data-residency
expectations~\cite{aiact2024} motivate on-premises models such as
Phi-4~\cite{phi42024}. Auditable evidence relies on RFC~3161
timestamps~\cite{rfc3161} and hash-chained custody in the tradition of
linked timestamping~\cite{haber1991timestamp}, and the
Cybersecurity Act~\cite{csact2019} anticipates machine-verifiable
certificates. Table~\ref{tab:peers} positions \sys: each alternative
covers a single column well, while \sys combines the three pillars on
a shared evidence layer with regulator-aligned mappings and an
on-premises RAG advisor.

\begin{table}[t]
\centering
\caption{Positioning of \sys against representative open-source
alternatives per pillar. \checkmark{} = native, $\circ$ = partial,
\,$-$\, = none.}
\label{tab:peers}
\renewcommand{\arraystretch}{1.0}
\setlength{\tabcolsep}{3pt}
\footnotesize
\begin{tabular}{@{}p{0.46\columnwidth}ccccc@{}}
\toprule
Capability & \rotatebox{75}{Eramba} & \rotatebox{75}{SimpleRisk}
           & \rotatebox{75}{MISP} & \rotatebox{75}{TheHive}
           & \rotatebox{75}{\sys} \\
\midrule
NIS2/CRA pre-mapped frameworks & $-$ & $-$ & $-$ & $-$ & \checkmark \\
Integrated scanner orchestration & $-$ & $-$ & $-$ & $\circ$ & \checkmark \\
MITRE ATT\&CK tactic mapping & $-$ & $-$ & \checkmark & \checkmark & \checkmark \\
STIX/TAXII CTI exchange & $-$ & $-$ & \checkmark & $\circ$ & \checkmark \\
On-premises AI compliance advisor & $-$ & $-$ & $-$ & $-$ & \checkmark \\
Hash-chained evidence + RFC\,3161 & $\circ$ & $-$ & $-$ & $\circ$ & \checkmark \\
ENISA-aligned cross-border IR & $-$ & $-$ & $-$ & $\circ$ & \checkmark \\
Cyber-Range on same back-end & $-$ & $-$ & $-$ & $-$ & \checkmark \\
\bottomrule
\end{tabular}
\end{table}

\section{User Requirements and Analysis}
\label{sec:reqs}

\subsection{Methodology and requirements}
\label{sec:reqs:method}

Kick-off workshops mapped stakeholder personas (Compliance Officer,
SOC analyst, external auditor, national CSIRT, pilot stakeholder,
system administrator) to consumed services. User stories became
functional and non-functional requirements
(\texttt{FR-<NNN>}/\texttt{NFR-<NNN>}), traced to the
Section~\ref{sec:intro} objectives and prioritised with MoSCoW; data
sources were declared per service and an initial KPI list cross-matched
with the use cases. Foundation requirements are multi-tenant isolation
and authentication (JSON Web Tokens (JWT), role-based access control
(RBAC), single sign-on (SSO), magic links); per-pillar requirements
cover multi-framework compliance with gap analysis, scanner
integration and cross-border CTI sharing, and a chain-of-custody audit
trail, each traced to a target in Table~\ref{tab:kpis}.

\subsection{Pilots and use cases}
\label{sec:reqs:pilots}

Five sectoral use cases drive validation, selected for
\emph{regulatory coverage} (major NIS2/CRA/GDPR obligation strands),
\emph{role diversity} (Compliance Officer, SOC analyst, external
auditor, CSIRT, law-enforcement agency (LEA)) and
\emph{cross-border interaction}
(notification, authority directories, multi-tenant residency). Each
has a baseline, a production-like operations phase and a controlled
attack-and-respond exercise (phishing and OAuth abuse, initial
foothold, lateral movement, data exfiltration, containment), scored on
the early-warning, notification, final-report and evidence-pack
artefacts the regulatory clocks demand (NIS2 Art.~23 at
24\,h/72\,h/one month, GDPR Art.~33 at 72\,h).

\emph{UC1 -- Energy / critical infrastructure}: NIS2-aligned
cross-border cybersecurity for power-grid operators. With OT in scope,
the profile is conservative: passive
techniques run on live OT, while active scanners (Nmap, OWASP ZAP) are
restricted to the IT periphery and sanitised lab replicas, never
production OT. It adds CTI exchange between national CSIRTs and energy
stakeholders and ENISA-conformant notification.

\emph{UC2 -- SME cybersecurity service provider}: managed-security
workflow for an SME delivering SOC, vulnerability-assessment (VA) and
pentest services to
enterprise clients: cross-client multi-tenancy, ISO/IEC~27001:2022
readiness and the SME-facing NIS2/CRA flow with magic-link
external-auditor review.

\emph{UC3 -- Public-sector cybersecurity authority} (national
CSIRT): regulatory monitor and notification across supervised
entities, with framework snapshots applied as NIS2 implementing acts
evolve.

\emph{UC4 -- Law enforcement}: forensic toolkit with
cross-jurisdictional digital-evidence collection, hash-chained
custody, RFC~3161 timestamps and lawful exchange under
LED~\cite{led2016} (and GDPR for personal data of
non-suspects~\cite{gdpr2016}), targeting ransomware, phishing and
misinformation reachable through the dark-web monitor and aligned
with Directive 2013/40/EU~\cite{dir2013_40}.

\emph{UC5 -- Smart-home / IoT manufacturer}: CRA path with
CE-marking checklists, SAST/SCA/SBOM ingestion from the product
pipeline, NVD/EUVD enrichment with CISA-KEV prioritisation,
conformity-assessment evidence packs and regulatory certificate
submission. The use cases overlap on suppliers so one supply-chain
event runs end-to-end from CRA-side manufacturer evidence to
NIS2-side essential-entity notification to the LEA evidence pack.

Platform-level targets are listed in Table~\ref{tab:kpis}, each
cross-mapped to the functional requirements (FR-IDs). All five use cases
are also exercised as Cyber-Range scenarios on a sanitised back-end
replica, where copied data is de-identified and operational tenants
are unreachable.

\begin{table}[t]
\centering
\caption{Platform-level pilot targets, by pillar, mapped to the
functional requirements (FR-IDs).}
\label{tab:kpis}
\renewcommand{\arraystretch}{1.0}
\setlength{\tabcolsep}{3pt}
\footnotesize
\begin{tabular}{@{}p{0.04\columnwidth}p{0.48\columnwidth}p{0.26\columnwidth}p{0.14\columnwidth}@{}}
\toprule
ID & Platform target & Target value & FRs \\
\midrule
\multicolumn{4}{l}{\emph{Cybersecurity pillar}}\\
K1 & Scanners integrated under one orchestrator
   & $\geq 5$ (Nmap, OSV, OWASP ZAP, Semgrep, Syft)
   & FR-009 \\
K2 & Threat-intelligence feeds synchronised
   & $\geq 4$ (CISA KEV, EUVD, MITRE ATT\&CK, NVD)
   & FR-013 \\
K3 & Dark-web engines under policy-gated allow-list
   & $\geq 20$
   & FR-014 \\
K4 & NIS2 early-warning produced
   & within 24~h
   & FR-004, FR-013 \\
K5 & NIS2 incident-notification produced
   & within 72~h
   & FR-004, FR-013 \\
\midrule
\multicolumn{4}{l}{\emph{Compliance pillar}}\\
K6 & Pre-seeded compliance frameworks
   & $\geq 5$ (CRA, GDPR, ISO/IEC 27001, NIS2, NIST CSF)
   & FR-003 \\
K7 & RAG retrieval latency, AI Assistant top-$k$
   & $\leq 2$~s end-to-end
   & FR-008 \\
K8 & Verifiable compliance certificates in pilots
   & $\geq 2$, SHA-256, 1-yr validity
   & FR-003, FR-011 \\
K9 & Framework snapshot \& revert under reg.\ update
   & $\geq 1$ per pilot
   & FR-003, FR-015 \\
\midrule
\multicolumn{4}{l}{\emph{Forensics pillar}}\\
K10 & Evidence items in hash-chained chain of custody
   & $\geq 50$ per pilot
   & FR-011 \\
K11 & RFC~3161 trusted timestamps on evidence
   & 100\% of admissible items
   & FR-011 \\
K12 & Cross-jurisdictional evidence packs re-verified by partner
   & $\geq 1$ per pilot
   & FR-011 \\
\midrule
\multicolumn{4}{l}{\emph{Training / Cyber-Range}}\\
K13 & Trainees onboarded via Cyber-Range exercises
   & $\geq 30$ across SOC, CSIRT, LEA, CO roles
   & FR-001, FR-002 \\
\bottomrule
\end{tabular}
\end{table}

\section{Architecture}
\label{sec:arch}

\sys is a modular microservice platform aligned with the ENISA
SOC/CSIRT reference architecture~\cite{enisa2023} and AI Act
transparency expectations~\cite{aiact2024}: REST APIs, horizontal
scaling, and security-by-design (least privilege, encrypted
transport, hash-chained evidence, trusted timestamps).

\subsection{Integration challenges and the shared evidence model}
\label{sec:arch:integ}

Each pillar is served in isolation by mature tooling. The difficulty
motivating \sys is not building any one of them but making a
\emph{single} artefact valid at once under three regulatory regimes
whose data models, obligations and timelines were never designed to
agree; integration that merely forwards reports between tools
reproduces the silos and leaves the operator to reconcile them by
hand. Five tensions make it non-trivial, and the shared hash-chained
evidence layer resolves each.

\begin{enumerate}
\item \emph{One artefact, three semantics.} A finding, CTI indicator
      or incident event must be at once a STIX/TAXII indicator carrying
      TLP and ATT\&CK context, a compliance evidence item bound to a
      Framework\,$\to$\,Objective, and a forensic custody record. These
      models differ in identity, schema and lifecycle, so \sys promotes
      every datum to one \texttt{Evidence} record with typed per-pillar
      projections rather than three copies that drift.
\item \emph{Conflicting obligations on the same datum.} GDPR and LED
      data minimisation pull toward redaction and short retention ---
      and privacy guarantees may not even compose across sequential
      releases of the same data~\cite{kekatos2026privacy} --- while
      forensic admissibility and CRA conformity pull toward
      complete, immutable records. The layer separates the integrity
      anchor (a SHA-256 hash and RFC~3161 timestamp~\cite{rfc3161} at
      intake) from the content (redacted of personally identifiable
      information (PII), minimised, re-fetchable only under
      authorisation), so a record stays verifiable after lawful
      minimisation.
\item \emph{One event, several regulatory clocks.} One incident
      starts the NIS2 24\,h, 72\,h and one-month clocks, the GDPR
      72\,h clock and the CRA vulnerability-handling clock, each
      anchored to a differently defined
      trigger~\cite{enisaincidentnotif2022}. Binding one timeline to
      several deadlines and dispatching to the right Member-State
      authority is a workflow-correctness problem, not a per-tool
      feature.
\item \emph{AI output as auditable evidence.} When the
      retrieval-augmented advisor drafts a notification or roadmap,
      that artefact may itself enter the chain. Admissibility requires
      binding model identity, retrieved sources and the human-review
      decision into the custody record, as the AI Act's transparency
      duties require~\cite{aiact2024}.
\item \emph{Evidence that crosses trust boundaries.} Tenant isolation
      and cross-border sharing pull in opposite directions: a partner
      CSIRT or competent authority must re-verify an exported pack
      without access to \sys, yet TLP scope, the GDPR/LED legal basis
      and residency constraints must travel with it. \sys attaches
      these as explicit export metadata so integrity and lawfulness
      are checkable offline.
\end{enumerate}

\noindent Fig.~\ref{fig:platform-overview} shows how the shared
evidence layer, with an ATT\&CK-anchored cooperation model, lets one
pillar's findings feed another's outcome.

\subsection{Threat model}
\label{sec:arch:threat}

\paragraph{Adversaries and surfaces} \sys considers three adversary
classes --- financially-motivated cybercriminals,
geopolitically-motivated actors and opportunistic exploiters of
unpatched products --- plus insider misuse. In-scope surfaces are the
public API, external-feed ingestion, cross-border exchange
(STIX/TAXII, evidence packs) and AI components (prompt poisoning,
regulated-data leakage); physical-layer attacks, prosecution-grade
image acquisition and upstream-scanner robustness are out of scope.

\paragraph{Mitigations} The three pillars provide complementary
responses, each mapped to relevant MITRE ATT\&CK~\cite{mitreattack}
tactics. Cybercriminals targeting SMEs and critical infrastructure
(initial access through lateral movement to ransomware impact) are met
with EU-aligned scanning under per-organisation policies,
KEV-prioritised findings and ENISA-aligned notification;
state-sponsored actors (resource development, persistence,
exfiltration) with dark-web monitoring, cross-border CTI sharing and
the forensic chain of custody; opportunistic exploiters of unpatched
products (exploitation, command-and-control) through the CRA
Asset\,$\to$\,Risk\,$\to$\,Control map, SBOM and patch management; and
insider misuse (credential access, collection) by RBAC and a
chain-of-custody audit trail. Across all four, hash-chained evidence
with RFC~3161, framework snapshots, intake-time PII redaction and
re-verifiable export keep the response auditable.

\subsection{System overview}
\label{sec:arch:overview}

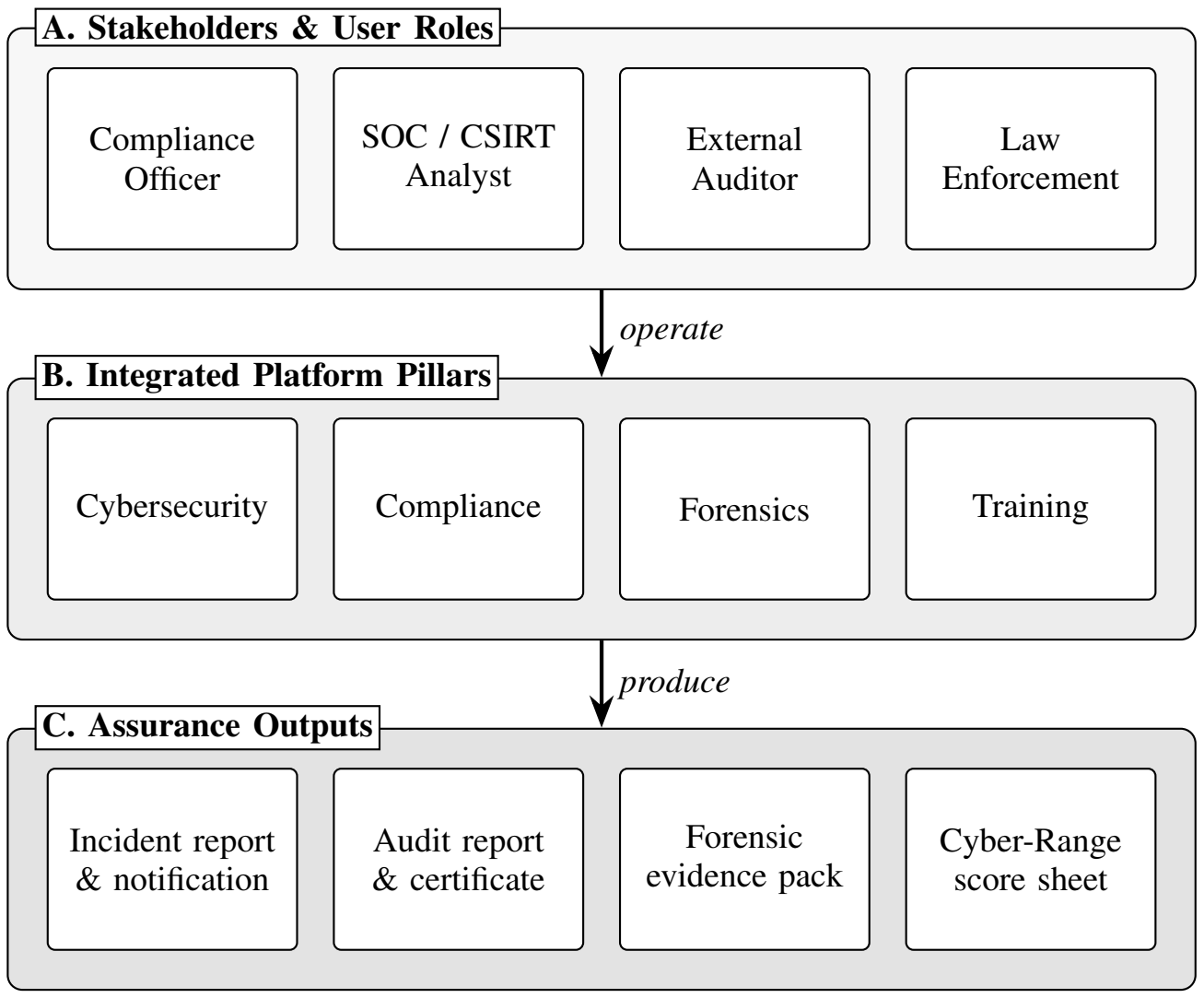
\begin{figure}[t]
\centering
\begin{tikzpicture}[
  font=\scriptsize,
  bx/.style={draw, rounded corners=1.5pt, fill=white, align=center,
             text width=16.5mm, minimum height=13mm, inner sep=2pt},
  varr/.style={-Stealth, thick},
  tab/.style={draw, fill=white, font=\scriptsize\bfseries, inner sep=1.3pt},
  cont/.style={draw, rounded corners=3pt, inner sep=2.8mm}
]
\node[bx] (a1) {Compliance Officer};
\node[bx, right=2.5mm of a1] (a2) {SOC / CSIRT Analyst};
\node[bx, right=2.5mm of a2] (a3) {External Auditor};
\node[bx, right=2.5mm of a3] (a4) {Law Enforcement};
\node[bx, below=12mm of a1] (p1) {Cybersecurity};
\node[bx, right=2.5mm of p1] (p2) {Compliance};
\node[bx, right=2.5mm of p2] (p3) {Forensics};
\node[bx, right=2.5mm of p3] (p4) {Training};
\node[bx, below=12mm of p1] (o1) {Incident report \& notification};
\node[bx, right=2.5mm of o1] (o2) {Audit report \& certificate};
\node[bx, right=2.5mm of o2] (o3) {Forensic evidence pack};
\node[bx, right=2.5mm of o3] (o4) {Cyber-Range score sheet};
\begin{scope}[on background layer]
  \node[cont, fill=black!3,  fit=(a1)(a4)] (boxA) {};
  \node[cont, fill=black!7,  fit=(p1)(p4)] (boxB) {};
  \node[cont, fill=black!11, fit=(o1)(o4)] (boxC) {};
\end{scope}
\node[tab, anchor=west] at ([xshift=2mm]boxA.north west) {A.\ Stakeholders \& User Roles};
\node[tab, anchor=west] at ([xshift=2mm]boxB.north west) {B.\ Integrated Platform Pillars};
\node[tab, anchor=west] at ([xshift=2mm]boxC.north west) {C.\ Assurance Outputs};
\draw[varr] (boxA.south) -- node[right, font=\scriptsize\itshape] {operate} (boxB.north);
\draw[varr] (boxB.south) -- node[right, font=\scriptsize\itshape] {produce} (boxC.north);
\end{tikzpicture}
\caption{\sys platform overview: end-user roles (tier~A) operate the
three platform pillars plus the Cyber-Range (tier~B), which produce
the audit-grade outcomes (tier~C), all bound to the shared
hash-chained evidence layer.}
\label{fig:platform-overview}
\end{figure}

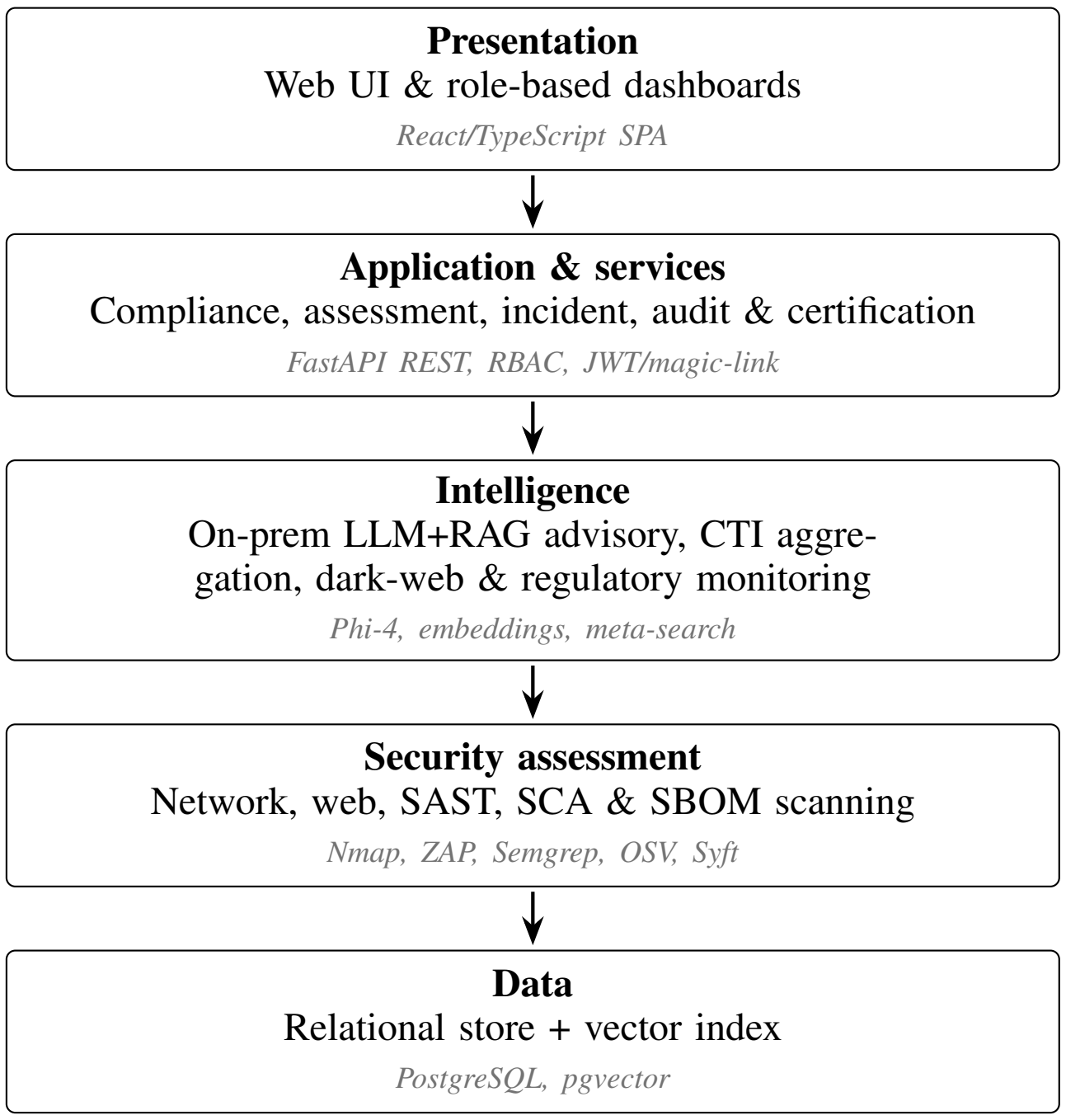
\begin{figure}[t]
\centering
\begin{tikzpicture}[
  font=\small,
  tier/.style={draw, rounded corners=2pt, minimum height=13mm,
               text width=82mm, align=center, inner sep=3.5pt, fill=white},
  arrow/.style={-Stealth, thick, shorten >=1pt, shorten <=1pt}
]
\node[tier] (t1)
  {\textbf{Presentation}\\Web UI \& role-based dashboards\\[1pt]
   {\ex React/TypeScript SPA}};
\node[tier, below=5mm of t1] (t2)
  {\textbf{Application \& services}\\Compliance, assessment, incident,
   audit \& certification\\[1pt]{\ex FastAPI REST, RBAC, JWT/magic-link}};
\node[tier, below=5mm of t2] (t3)
  {\textbf{Intelligence}\\On-prem LLM+RAG advisory, CTI aggregation,
   dark-web \& regulatory monitoring\\[1pt]
   {\ex Phi-4, embeddings, meta-search}};
\node[tier, below=5mm of t3] (t4)
  {\textbf{Security assessment}\\Network, web, SAST, SCA \& SBOM
   scanning\\[1pt]{\ex Nmap, ZAP, Semgrep, OSV, Syft}};
\node[tier, below=5mm of t4] (t5)
  {\textbf{Data}\\Relational store + vector index\\[1pt]
   {\ex PostgreSQL, pgvector}};
\draw[arrow] (t1) -- (t2);
\draw[arrow] (t2) -- (t3);
\draw[arrow] (t3) -- (t4);
\draw[arrow] (t4) -- (t5);
\end{tikzpicture}
\caption{\sys reference architecture: five capability tiers (example
implementations in italic), with authentication, audit logging and
trusted timestamping cross-cutting all tiers. External sources feed
the intelligence and assessment tiers: CISA KEV, EUVD, NVD, MITRE
ATT\&CK, FreeTSA (RFC~3161) and Tor. Compliance services host the
regulatory frameworks (CRA, GDPR, ISO/IEC~27001, NIS2, NIST~CSF) as
Framework\,$\to$\,Chapter\,$\to$\,Objective.}
\label{fig:arch}
\end{figure}

\sys is a Docker-orchestrated five-tier microservice stack
(Fig.~\ref{fig:arch}), with authentication, audit logging and trusted
timestamping cutting across all tiers and external feeds entering
through the backend; notification endpoints form a configurable
authority directory. End-to-end, the platform connects the user roles
to audit-grade outcomes via the three pillars
(Fig.~\ref{fig:platform-overview}).

\section{Toolkits and Platform Services}
\label{sec:toolkits}

The platform groups its services into three toolkits (one per
pillar), with Cyber-Range scenarios layered on top, all sharing the
evidence layer of Section~\ref{sec:arch:integ}.

\subsection{Cybersecurity Toolkit}
\label{sec:tk-cyber}

Scanners, the CTI and dark-web services, the regulatory monitor and
the incident workflow run under per-organisation policies; every
finding is wrapped in the common hash-chained custody envelope, so the
same artefact is citable in a conformity pack and a partner CSIRT
exchange.

\paragraph{Scanning and CTI}
The scan orchestrator dispatches Nmap, OWASP ZAP, Semgrep, OSV and
Syft over HTTP, normalises and deduplicates findings, enriches with
NVD/EUVD and CISA-KEV~\cite{cisakev}, and maps them onto MITRE
ATT\&CK~\cite{mitreattack} in the CTI service. Indicators are shared
across borders via STIX/TAXII~\cite{stixtaxii} envelopes with TLP
markings, addressing long-standing CTI-sharing
challenges~\cite{wagner2019cti}; an AI layer prioritises by recipient
asset profile to reduce alert fatigue.

\paragraph{Dark-web and regulatory monitor}
A Tor-fronted scanner queries an allow-listed engine set under
per-organisation policies. Hits are classified, hashed and
PII-redacted; the store keeps only evidence-grade pointers (URL,
content hash, classifier labels), full content fetched transiently on
a \emph{minimum-necessary} basis. A SearXNG meta-search drives the
regulatory monitor: an LLM proposes framework deltas that an
administrator accepts, triggering a snapshot.

\paragraph{Incident management}
The module implements an ENISA-aligned workflow (early warning,
notification, final report), a forensic timeline linking evidence
to events, and submissions to national CSIRTs.

\paragraph{End-to-end walk-through}
One step of the attack-and-respond (Section~\ref{sec:reqs:pilots}): an
OWASP-ZAP scan returns a critical, KEV-flagged remote-code-execution
flaw (Common Vulnerability Scoring System (CVSS)~9.1) on a
client-facing portal. The CTI service normalises it, maps it to the
Initial Access tactic via technique T1190~\cite{mitreattack} and shares
it with partner CSIRTs as a STIX/TAXII bundle under a restricted TLP
marking. The compliance engine surfaces it against the NIS2
notification objective and the CRA essential requirements (Annex~I),
and the AI Compliance Agent drafts the early-warning notification and a
remediation roadmap with due dates. The forensic layer enrols the scan
output, vulnerability record and AI draft into the engagement's hash-chained
custody with RFC~3161 timestamps; the same trace becomes a Cyber-Range
exercise scored on notification timeliness and ATT\&CK mapping accuracy.

\subsection{Compliance \& Assurance Toolkit}
\label{sec:tk-compliance}

\paragraph{Framework management}
Frameworks (CRA, GDPR, ISO/IEC 27001, NIS2, NIST CSF) sit in a
Framework\,$\to$\,Chapter\,$\to$\,Objective hierarchy; each
Objective carries a status and links to policies, risks, evidence
and answers. A snapshot mechanism applies regulatory updates while
preserving revert to a prior baseline.

\paragraph{Compliance engine and assessments}
Users create assessments against a chosen framework; the engine
computes gaps from answer records and surfaces them on role-specific
dashboards. The Compliance Chain Visualiser renders the
Risk\,$\to$\,Policy\,$\to$\,Objective\,$\to$\,Framework chain, so a gap
on, e.g., the NIS2 notification objective is traceable to the risk and
policy that own it.

\paragraph{Risk and asset management}
A risk register (likelihood, residual risk, severity, status, product
linkage) is paired with a CRA-aligned product catalogue (economic
operator, product type, criticality with confidentiality, integrity,
availability (CIA) ratings), and CE-marking
checklists are auto-generated. A curated \texttt{cra\_connections} map
links every CRA Annex~I/Ch.~II objective to its risks, controls and
policies, so any CRA-scoped asset surfaces its
Asset\,$\to$\,Risk\,$\to$\,Control\,$\to$\,Policy chain.

\paragraph{AI Assistant and Compliance Agent}
The \emph{AI Assistant} is RAG over a pgvector index of seeded
framework corpora and organisation policies on a local
Phi-4~\cite{phi42024,lewis2020rag}; every response carries the
retrieved citations that ground it~\cite{aiact2024}. The
\emph{Compliance Agent} reuses that retrieval with the risk register,
recent scans and open incidents to produce a structured roadmap
attached to the originating objective, with provider adapters for
hosted models where data-residency permits.

\paragraph{Evidence, audit and certification}
Every artefact (scan findings, timeline entries, policy acceptances,
AI remediations, generated PDFs) is wrapped as an \texttt{Evidence}
record (SHA-256 hash, predecessor handle, RFC~3161
timestamp~\cite{rfc3161}, framework binding, optional legal-hold).
Certificates are SHA-256-verifiable with a configurable validity
window and reach competent authorities through pre-configured
endpoints; auditors connect via magic-link only to invited
engagements, with reads and annotations also chained.

\subsection{Cross-Jurisdictional Forensic Toolkit}

\paragraph{Forensic timeline}
Every incident carries a timeline linking events to artefacts
(scanner findings, CTI indicators, responder logs), filtered by actor,
asset and MITRE ATT\&CK technique and exported as an evidence-grade
PDF. Each event card carries its actor, asset, technique and content
hash, and the export manifest records the chain head, RFC~3161
timestamp and TLP marking, making the shared evidence layer of
Section~\ref{sec:arch:integ} directly visible; the same custody
envelope is reused for cross-jurisdictional exchange.

\paragraph{Cross-jurisdictional exchange}
Evidence packs export with their custody chain and RFC~3161 tokens
so a partner can re-verify integrity without access to \sys.
Transfer is lawful by design: personal data is enumerated under GDPR
Art.~30 with a stated legal basis (Art.~6)~\cite{gdpr2016}; exports by
competent authorities processing for prevention, investigation or
prosecution are governed by the LED~\cite{led2016} (purpose limitation,
categorisation of data subjects), aligned with Directive
2013/40/EU~\cite{dir2013_40} where applicable.
Law-enforcement access uses magic-link engagements with an explicit
authorisation artefact attached.

\section{Discussion}
\label{sec:discussion}

Section~\ref{sec:arch:integ} set out why unifying the three pillars is
non-trivial and how the shared evidence layer resolves it; we close on
the open questions and limitations.

\paragraph{Overlapping frameworks}
NIS2, the CRA and GDPR share controls but differ in obligation
language and evidence expectations; \sys runs them as parallel
frameworks with explicit cross-mappings. Whether these mappings reduce
or merely shift manual effort is the empirical question the pilots
will answer.

\paragraph{AI advisor trustworthiness}
RAG grounds the advisor but hallucination remains possible; citation
transparency and human-in-the-loop review are essential, and
depending on deployment such assistance may trigger additional AI Act
transparency obligations~\cite{aiact2024}. Calibrated abstention and
post-hoc consistency checks are open work.

\paragraph{Cross-border interop and limitations}
Member-State CSIRTs implement NIS2 differently, so notification
endpoints are a configurable authority directory pending
machine-readable APIs. The forensic toolkit \emph{supports}
admissibility (hash-chained custody, RFC~3161, legal hold,
re-verifiable export), but the pilot's FreeTSA is not eIDAS-qualified,
so evidence feeds national procedures rather than substituting for
them. Replacing FreeTSA with an eIDAS-qualified trust service provider is on the
road-map; pilot KPIs, AI accuracy and latency arrive in later
deliverables.

\section{Conclusions}
\label{sec:conclusions}

This paper introduced \sys, an EU-funded AI-powered platform
integrating cross-border cybersecurity, compliance and forensics under
one architecture and a shared evidence layer, with a Cyber-Range on a
sanitised back-end replica. The contribution is the integration:
hash-chained custody with RFC~3161 timestamps binds the three pillars
so a CTI indicator, an assessment answer and a forensic artefact carry
the same rules, and one supply-chain incident runs end-to-end across
the five use cases. Next-phase
work will report pilot KPIs, mature AI calibration and confidence
reporting, replace FreeTSA with an eIDAS-qualified trust service provider, and move
certification towards machine-verifiable artefacts aligned with the
Cybersecurity Act~\cite{csact2019} scheme.

\section*{Acknowledgements}
This work has received funding from the European Union's Digital
Europe Programme under grant agreement No 101249702 (CYBER-BRIDGE).

\bibliographystyle{IEEEtran}
\bibliography{references}

\end{document}